\documentclass[apj]{emulateapj}
\usepackage{apjfonts}

\slugcomment{Accepted for publication in ApJ Letters}

\shorttitle{Morphological evolution of galaxies to $z\sim 2$.}
\shortauthors{Szomoru et al.}
\begin{document}

\title{Morphological evolution of galaxies from ultradeep HST WFC3
  imaging: \\ the Hubble sequence at $z\sim 2$.\altaffilmark{1}}

\author{
Daniel Szomoru\altaffilmark{2},
Marijn Franx\altaffilmark{2},
Rychard~J. Bouwens\altaffilmark{2},
Pieter~G.~van Dokkum\altaffilmark{3},
Ivo Labb\'e\altaffilmark{4},
Garth~D. Illingworth\altaffilmark{5},
Michele Trenti\altaffilmark{6}
}

\altaffiltext{1}
{Based on observations with the NASA/ESA {\em Hubble Space
Telescope}, obtained at the Space Telescope Science Institute, which
is operated by AURA, Inc., under NASA contract NAS 5--26555.}
\altaffiltext{2}
{Leiden Observatory, Leiden University, P.O. Box 9513, 2300 RA Leiden, The Netherlands}
\altaffiltext{3}
{Department of Astronomy, Yale University, New Haven, CT 06520-8101, USA}
\altaffiltext{4}
{Carnegie Observatories, Pasadena, CA 91101, USA}
\altaffiltext{5}
{UCO/Lick Observatory, University of California, Santa Cruz, CA 95064, USA}
\altaffiltext{6}
{Center for Astrophysics and Space Astronomy, University of Colorado, 389-UCB, Boulder, CO 80309, USA}

\begin{abstract}
We use ultradeep HST WFC3/IR imaging of the HUDF to investigate the
rest-frame optical morphologies of a mass-selected sample of galaxies
at $z\sim 2$. We find a large variety of galaxy morphologies, ranging
from large, blue, disk-like galaxies to compact, red, early-type
galaxies. We derive rest-frame $u-g$ color profiles for these galaxies
and show that most $z\sim 2$ galaxies in our sample have negative
color gradients such that their cores are red. Although these color
gradients may partly be caused by radial variations in dust content,
they point to the existence of older stellar populations in the
centers of $z\sim 2$ galaxies. This result is consistent with an
``inside-out'' scenario of galaxy growth. We find that the median
color gradient is fairly constant with redshift:
$(\Delta(u-g_\mathrm{rest})/\Delta(\log{r}))_\mathrm{median} = -0.46$,
$-0.44$ and $-0.49$ for $z\sim 2$, $z\sim 1$ and $z=0$,
respectively. Using structural parameters derived from surface
brightness profiles we confirm that at $z\sim 2$ galaxy morphology
correlates well with specific star formation rate. At the same mass,
star forming galaxies have larger effective radii, bluer rest-frame
$u-g$ colors and lower S\'ersic indices than quiescent galaxies. These
correlations are very similar to those at lower redshift, suggesting
that the relations that give rise to the Hubble sequence at $z=0$ are
already in place for massive galaxies at this early epoch.
\end{abstract}


\section{Introduction}

\begin{figure*}
\epsscale{1.2}
\plotone{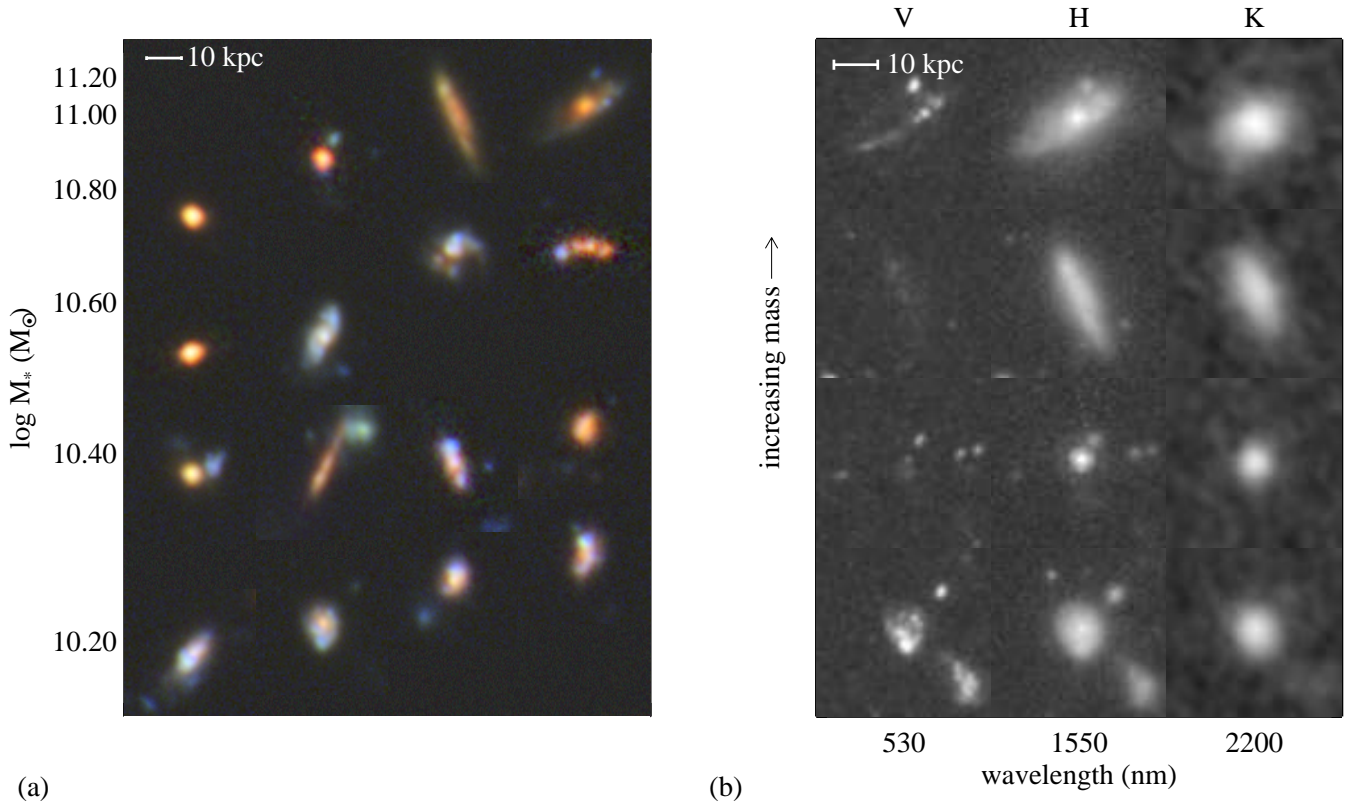}
\caption{(\textit{a}) Morphologies of $z\sim 2$ galaxies in the
  HUDF. The blue, green and red color channels are composed of
  PSF-matched HST/ACS $i_{775}$, HST/WFC3 $Y_{105}$ and HST/WFC3
  $H_{160}$ images, respectively. Mass increases upwards. The galaxies
  exhibit a very large range in morphologies: from very compact,
  nearly unresolved spheroids with red colors, to large, extended
  disk-like galaxies with blue colors.  Most of the galaxies have a
  well-defined center, which is usually redder than the outer parts of
  the galaxy. (\textit{b}) Four massive galaxies at $z\sim 2$. From
  left to right, the galaxies are shown as observed in the HST/ACS
  $V_{606}$ band, the HST/WFC3 $H_{160}$ band, and the (ground-based)
  $K_s$ band. The difference between the (PSF matched) $V_{606}$ band
  and $H_{160}$ band images is large: whereas the galaxies exhibit
  very regular morphologies and have well-defined centers in the
  $H_{160}$ band, they are very clumpy and irregular in the $V_{606}$
  band, and in some cases are nearly undetected. These complex,
  wavelength-dependent morphologies suggest that $z\sim 2$ galaxies
  may be composed of multiple components with very different stellar
  ages.}\label{fig:z2galaxies}
\end{figure*}

Since the description of the Hubble sequence (e.g., \citealt{hub26};
\citealt{vau59}; \citealt{san81}) we have learned that position along
the Hubble sequence correlates with parameters like color, stellar
age, and gas fraction (e.g., \citealt{rob94}). However, morphologies
by themselves provide very limited information, as they are scale free
and do not include physical parameters such as surface brightnesses,
sizes, luminosities and masses. Using the Sloan Digital Sky Survey
(SDSS, \citealt{yor00}), \cite{kau03} showed that the main parameter
driving galaxy properties is stellar mass. High mass galaxies are
generally red, have old stellar populations and low specific star
formation rates (SSFRs), while low mass galaxies are generally blue,
have young stellar populations and high SSFRs.

A key question is what the structure was of the progenitors of low
redshift galaxies. Hubble Space Telescope studies out to redshift
$z\sim 1$ indicate that the morphological variation is comparable to
that at low redshift (e.g., \citealt{bel04}). More recent studies of
$z > 2$ galaxies using the HST NICMOS camera have yielded varying
results, largely due to different selection criteria. For example,
\cite{pap05} studied the rest-frame optical morphologies of a
flux-limited sample of galaxies at $z \approx 2.3$ and found that they
are generally irregular. \cite{tof05}, on the other hand, investigated
the rest-frame optical and UV morphologies of distant red galaxies
(DRGs) in the Hubble Ultra Deep Field (HUDF), and found both galaxies
with irregular morphologies and galaxies with smooth
morphologies. Additionally, they showed that the rest-frame optical
morphologies of these galaxies are much more regular and centrally
concentrated than the rest-frame UV morphologies.

With the advent of the Wide Field Camera 3 (WFC3), with its vastly
improved sensitivity and resolution compared to NICMOS, it has become
possible to analyze the rest-frame optical structure of high redshift
galaxies with an unprecedented level of detail. \cite{cam10} have used
data from the first year of observations of the HUDF and the Early
Release Science Field to classify the rest-frame UV and optical
morphologies of galaxies up to $z\sim 3.5$. These authors confirm
results by e.g. \cite{kri09}, who showed that massive galaxies at
$z\approx 2.3$ can be separated into two distinct classes: blue
star-forming galaxies with irregular morphologies on the one hand, and
red quiescent galaxies with smoother morphologies on the other.

In this Letter, we extend the previous results using the full two-year
ultradeep near-infrared (NIR) imaging of the HUDF taken with the HST
WFC3. These data are the deepest ever obtained in the NIR and make it
possible to analyze the morphologies, colors and structure of galaxies
to $z\sim 3$ in the rest-frame optical. Using the incredible
sensitivity and angular resolution of the WFC3 images we analyze the
rest-frame optical surface brightness profiles of a mass-selected
sample of galaxies at $z\sim 2$. We use these profiles to derive
structural parameters such as size and profile shape, and obtain
rest-frame color profiles. We study the correlations between these
parameters as a function of redshift in order to investigate the
Hubble sequence at different epochs in the history of the
Universe. Throughout the Letter, we assume a $\Lambda$CDM cosmology
with $\Omega_m = 0.3$, $\Omega_\Lambda = 0.7$ and $H_0 = 70$ km
s$^{-1}$ Mpc$^{-1}$.

\section{Morphologies at $z\sim 2$}

\begin{deluxetable*}{l l c c c c c c c c}
\tablewidth{0pt}
\tablecaption{Galaxy properties.\label{tab:table}}
\tablehead{
  \colhead{Source} & \colhead{ID$^\mathrm{a}$} & \colhead{mag$_{app}^\mathrm{b}$} & \colhead{$r_e^\mathrm{b}$} & \colhead{$n^\mathrm{b}$} & \colhead{$M_{stellar}^\mathrm{c}$} & \colhead{$u-g_{rest}$} & \colhead{log $\Sigma_{ave}^\mathrm{d}$} & \colhead{$z^\mathrm{c}$} \\
\colhead{} & \colhead{} & \colhead{(AB)} & \colhead{(kpc)} & \colhead{} & \colhead{($M_\odot$)} & \colhead{} & \colhead{($M_\odot$ kpc$^{-2}$)} & \colhead{}}
\startdata
HUDF09 & 3203 & 21.81 &  3.49 &  1.15 & 10.65 &  0.49 &  8.77 & 1.998 \\
HUDF09 & 3239 & 23.11 &  0.32 &  2.06 & 10.51 &  1.18 & 10.72 & 1.980 \\
HUDF09 & 3242 & 22.10 &  0.44 &  3.21 & 10.76 &  1.18 & 10.67 & 1.910 \\
HUDF09 & 3254 & 23.31 &  1.96 &  1.31 & 10.25 &  0.85 &  8.87 & 1.887* \\
HUDF09 & 3391 & 23.28 &  2.13 &  0.44 & 10.37 & -0.04 &  8.92 & 1.919* \\
HUDF09 & 3421 & 23.79 &  2.12 &  0.90 & 10.59 &  0.19 &  9.14 & 2.457* \\
HUDF09 & 3463 & 22.47 &  1.80 &  1.20 & 10.09 &  0.59 &  8.78 & 1.659* \\
HUDF09 & 3486 & 22.43 &  2.78 &  0.65 & 10.22 &  0.88 &  8.53 & 1.628* \\
HUDF09 & 3595 & 22.39 &  3.67 &  0.41 & 10.98 &  0.87 &  9.06 & 1.853* \\
HUDF09 & 3634 & 22.94 &  2.37 &  1.28 & 10.08 &  0.58 &  8.54 & 1.994 \\
HUDF09 & 3653 & 23.85 &  1.94 &  0.95 & 10.27 &  1.24 &  8.90 & 1.776* \\
HUDF09 & 3721 & 22.02 &  3.00 &  1.16 & 10.54 &  0.63 &  8.79 & 1.843 \\
HUDF09 & 3757 & 23.10 &  1.25 &  5.04 & 10.28 &  1.20 &  9.29 & 1.674* \\
HUDF09 & 3799 & 24.59 &  2.93 &  1.50 & 10.62 &  0.87 &  8.89 & 2.492* \\
HUDF09 & 6161 & 21.58 &  6.70 &  3.51 & 11.08 &  1.22 &  8.63 & 1.552 \\
HUDF09 & 6225 & 23.26 &  2.58 &  0.56 & 10.33 &  0.72 &  8.71 & 2.401* \\
HUDF09 & 6237 & 23.51 &  < 0.10 & > 10.00 & 10.72 &  1.30 & 14.40 & 1.965* \\
\nodata & \nodata & \nodata & \nodata & \nodata & \nodata & \nodata & \nodata & \nodata \\
\enddata
\tablecomments{This is a sample of the full table, which is available in the online version of the article (see Supporting Information).}
\tablenotetext{a}{IDs correspond to NYU-VAGC IDs \citep{bla05} for the SDSS galaxies, and FIREWORKS IDs \citep{wuy08} for the HUDF galaxies}
\tablenotetext{b}{These parameters are derived from the SDSS $g$ band imaging for the SDSS galaxies, and from the HST/WFC3 $H_{160}$ band imaging for the HUDF09 galaxies}
\tablenotetext{c}{Stellar masses and redshifts were obtained from the \cite{guo09} catalog for the SDSS galaxies, and the \cite{wuy08} catalog for the HUDF galaxies. All masses include a correction factor to account for the difference between the catalog magnitude and our measured magnitude.}
\tablenotetext{d}{$\Sigma_{ave}$ is defined as the average surface density within $r_e$}
\tablenotetext{*}{No spectroscopic redshifts are available for these galaxies; photometric redshifts are listed instead}
\end{deluxetable*}

We use the full two-year data taken with WFC3/IR of the HUDF, obtained
in 2009-2010 as part of the HUDF09 HST Treasury program (GO11563). It
consists of 24 orbits of $Y_{105}$ imaging, 34 orbits of $J_{125}$
imaging, and 53 orbits of $H_{160}$ imaging.  These images were
reduced using an adapted pipeline (\citealt{bou10},
\citealt{oes10}). The FWHM of the point-spread function (PSF) is
$\approx 0.16$ arcsec.  The images are combined with very deep
Advanced Camera for Surveys (ACS) images of the HUDF \citep{bec06} to
construct color images of the massive galaxies in the field.

We use the $K_s$-selected catalog of \cite{wuy08} to select galaxies
in the HUDF for which WFC3 imaging is available. This catalog combines
observations of the Chandra Deep Field South ranging from ground-based
$U$ band data to Spitzer $24 \mu$m data, and includes spectroscopic
redshifts where available, as well as photometric redshifts derived
using EAZY \citep{bra08}. Stellar masses were estimated from spectral
energy distribution fits to the full photometric data set (F\"orster
Schreiber et al., 2011 in preparation), assuming a Kroupa initial mass
function and the stellar population models of \cite{bru03}. To study
the morphological variation at $z\sim 2$, we select the 16 most
massive galaxies with $1.5 < z < 2.5$. These galaxies have stellar
masses between $1.2 \times 10^{10}$ M$_\odot$ to $1.3 \times 10^{11}$
M$_\odot$. Color images of these galaxies are shown in
Figure~\ref{fig:z2galaxies}a. A summary of their properties is given
in Table~\ref{tab:table}.

From Figure~\ref{fig:z2galaxies}a it is apparent that galaxies at
$z\sim 2$ show a large variation in morphology, size and color. One
can distinguish red, smooth, compact galaxies; blue galaxies with
disk-like structures, some even with apparent spiral arms; and other
star forming galaxies which appear more irregular. Most of the
galaxies have a well-defined, red center. This is further illustrated
in Figure~\ref{fig:z2galaxies}b, where galaxy morphology is shown as a
function of wavelength. Four massive galaxies are shown in the
observed HST/ACS $V_{606}$ band \citep{bec06}, HST/WFC3 $H_{160}$ band
and $K_s$ band (based on ground-based imaging by Labb\'e et al., in
preparation). The morphology of the sources differs strongly as a
function of wavelength. The rest-frame UV morphology is always more
clumpy and extended than the rest-frame optical morphology. In
addition, the rest-frame optical images show well-defined centers for
all our sources, whereas these are very often lacking in the
rest-frame UV images. This confirms that the mass distribution of
$z\sim 2$ galaxies is smoother and more centrally concentrated than
would be concluded from rest-frame UV imaging (see e.g.,
\citealt{lab03b}; \citealt{tof05}).

\section{Color gradients at $z\sim 2$}

\begin{figure*}
\epsscale{1.2}
\plotone{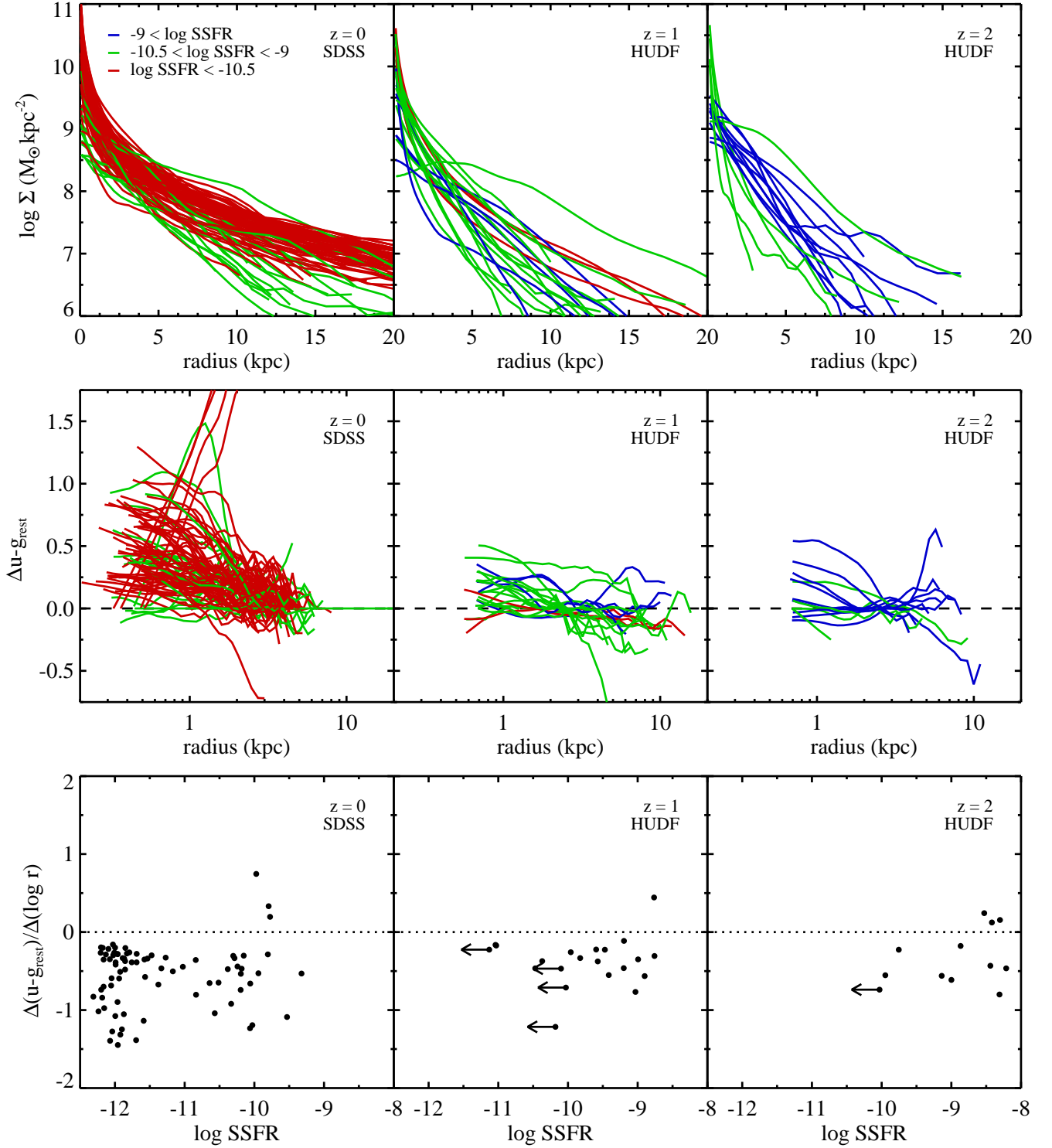}
\caption{\textit{Top panels:} surface density profiles of galaxies at
  $z = 0$, $z\sim 1$ and $z\sim 2$, color-coded according to specific
  star formation rate. The profiles are plotted up to the radius where
  uncertainties in the sky determination become significant. Surface
  density has been calculated from the average $M/L_H$ ($M/L_r$ for
  the SDSS galaxies) of the galaxy multiplied by the observed
  $H_{160}$ ($r_{625}$) band surface brightness profile, ignoring
  gradients in the mass-to-light ratio. It is clear that a large
  diversity of profiles exists, with some close to exponential
  profiles and others close to $r^{1/4}$ profiles. This diversity
  extends to $z\sim 2$. \textit{Middle panels:} rest-frame $u-g$ color
  profiles of the same galaxies. The profiles are normalized so that
  $\Delta u-g_{rest} = 0$ at $r = r_e$. The color profiles are plotted
  from the PSF HWHM ($\sim 0.6$ arcsec for the SDSS galaxies, $\sim
  0.08$ arcsec for the HUDF galaxies) to the radius where the errors
  in the flux measurement reach $20\%$. The color profiles show an
  overall trend of redder colors at smaller radii, in all redshift
  intervals. \textit{Bottom panels:} rest-frame $u-g$ color
  gradients. Arrows indicate upper limits. Most $z\sim 2$ galaxies
  have negative color gradients. Color gradients do not seem to evolve
  very strongly between $z\sim 2$ and $z=0$, for galaxies in the mass
  range considered here.}\label{fig:profiles}
\end{figure*}

In order to quantify the morphological properties of $z\sim 2$
galaxies we measure their surface brightness profiles. In contrast to
conventional model-fitting techniques where a simple model is used to
approximate the intrinsic surface brightness profile (e.g., using the
GALFIT package of \citealt{pen02}), we measure the actual profile
using the approach of \cite{szo10}. The intrinsic profile is derived
by fitting a S\'ersic model profile convolved with the PSF to the
observed flux, and then adding the residuals from this fit to the
unconvolved model profile. Effectively, the model profile is used to
deconvolve the majority of the observed flux, after which this
deconvolved profile is combined with the residuals to account for
deviations from the assumed model. \cite{szo10} have shown that it is
thus possible to accurately measure the true intrinsic profiles out to
large radii and very low surface brightness, even in cases where
galaxies comprise a bright compact bulge and a faint extended disk.

In addition to the $z\sim 2$ galaxies we measure the profiles of
galaxies at $z\sim 1$ and $z = 0$. The $z\sim 1$ sample is taken from
the same dataset as the $z\sim 2$ sample, whereas the $z=0$ galaxies
are taken from the \cite{guo09} SDSS catalog of central galaxies. The
galaxies are selected to lie in the same mass interval as the $z\sim
2$ sample: $1.2 \times 10^{10} M_\odot < M_{stellar} < 1.3 \times
10^{11} M_\odot$. The SDSS redshifts are required to be $z = 0.03 \pm
0.015$ and the $z\sim 1$ galaxies have $0.5 < z < 1.5$. This results
in a sample of 27 galaxies at $z\sim 1$ and 84 galaxies at
$z=0$. Stellar masses and SSFRs for the SDSS sample are obtained from
the MPA/JHU
datarelease\footnote[1]{http://www.mpa-garching.mpg.de/SDSS/} (see
\citealt{bri04}; \citealt{sal07} for details). A summary of the
properties of the $z\sim 1$ and $z=0$ galaxies is given in
Table~\ref{tab:table}.

In Figure~\ref{fig:profiles} we show surface density profiles,
$u-g_{rest}$ color profiles, and $u-g_{rest}$ color gradients for
galaxies in all three redshift bins. The surface density profiles are
obtained by multiplying the surface brightness profiles in the
$H_{160}$ band ($g$ band for the SDSS galaxies) with the galaxies'
average $M_{stellar}/L$ ratios in that band (where $M_{stellar}$ is
the total mass from the \citealt{wuy08} catalog (MPA/JHU catalog for
the SDSS galaxies) and $L$ is derived from the surface brightness
profiles). This approach ignores gradients in the $M_{stellar}/L$
ratios and is therefore not exact; however, it allows for a more
quantitative comparison between galaxies at different redshifts and
with different colors. The color gradients are derived from fits to
the $u-g$ profiles between the half-width at half-maximum (HWHM) of
the PSF ($\sim 0.6$ arcsec for the SDSS galaxies, $\sim 0.08$ arcsec
for the HUDF galaxies) and the radius where the errors in the flux
measurement reach 20\%.

The majority of $z\sim 2$ galaxies in our sample have negative color
gradients, which do not vary strongly with redshift:
$(\Delta(u-g_\mathrm{rest})/\Delta(\log{r}))_\mathrm{median} =
-0.47^{+0.20}_{-0.56}$, $-0.33^{+0.19}_{-0.23}$ and
$-0.46^{+0.58}_{-0.28}$ for $z\sim 2$, $z\sim 1$ and $z=0$,
respectively (where the errors give the 1-$\sigma$ interval around the
median). Thus, galaxy color gradients seem to be remarkably constant
with redshift, both for quiescent and star-forming galaxies. It should
be noted that this is not a comparison of low-redshift galaxies to
their high-redshift progenitors; the $z\sim 2$ galaxies are expected
to evolve into more massive galaxies at low redshift, due to mergers
and accretion. Since we use mass-limited samples, the $z\sim 2$
galaxies we consider will most likely fall outside of our mass-limits
at low redshift (see, e.g., \citealt{dok10}).

\section{The Hubble sequence from $z\sim 2$ to $z=0$}\label{sec:profiles}

\begin{figure*}
\epsscale{1.2}
\plotone{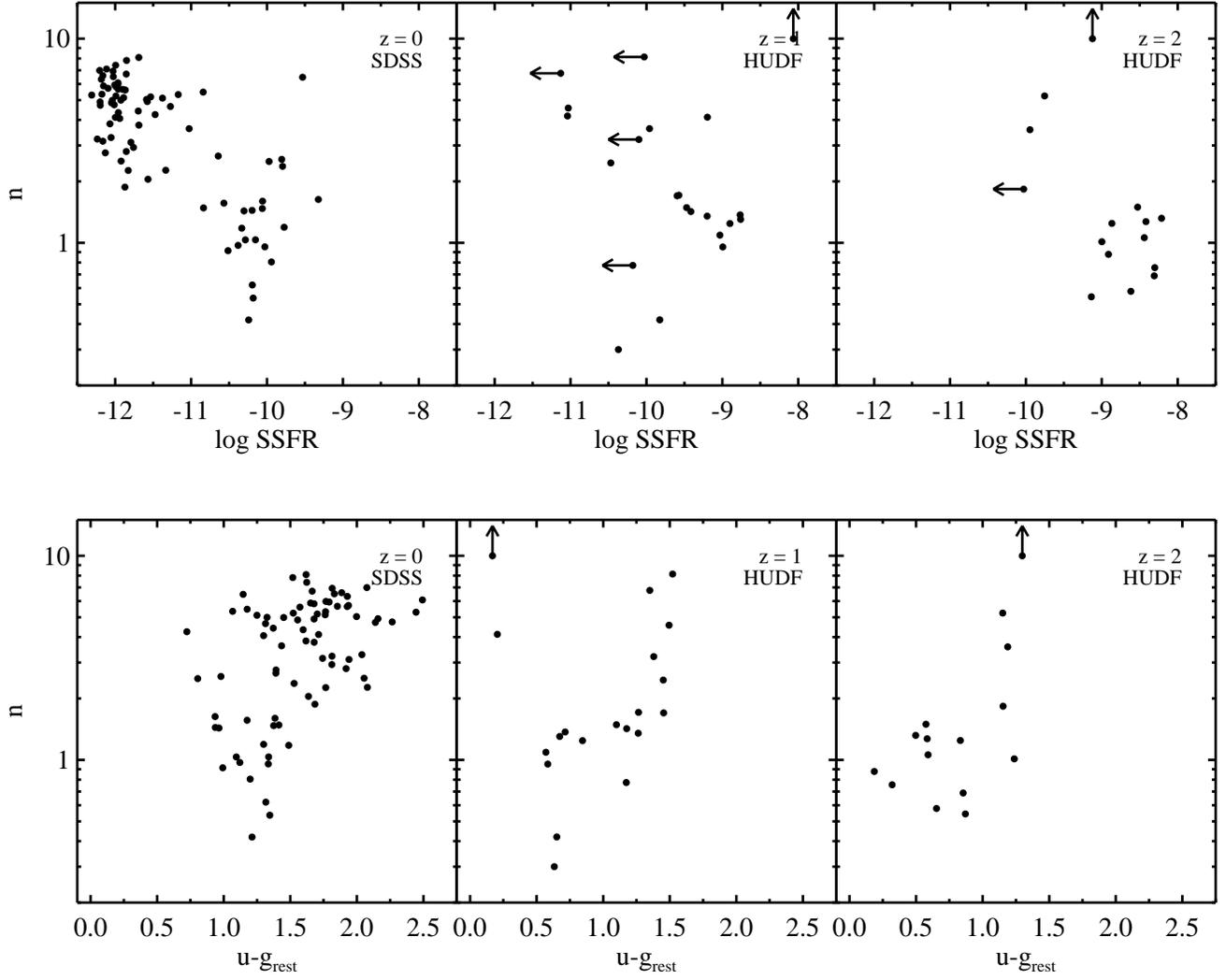}
\caption{S\'ersic index $n$ plotted against SSFR (\textit{top row})
  and $u-g_{rest}$ color (\textit{bottom row}) for galaxies with
  $1.2\times 10^{10} M_\odot < M_{stellar} < 1.3\times 10^{11}
  M_\odot$, at $z=0$, $z\sim 1$ and $z\sim 2$. Arrows indicate upper
  or lower limits. Profile shape and SSFR are anticorrelated in all
  three wavelength bins, whereas profile shape and color show a
  positive correlation. The relations between these morphological
  parameters are similar in all three redshift bins, although on
  average galaxies are bluer and have higher SSFRs and lower S\'ersic
  indices at high redshift. The similarity of the relations at $z=0$
  and $z\sim 2$ suggests that the Hubble sequence was already in place
  for massive galaxies at $z\sim 2$.}\label{fig:hubble}
\end{figure*}

Finally, we study the relations between structure, color and SSFR as a
function of redshift in Figure~\ref{fig:hubble}. In the top row we
show S\'ersic index against SSFR for the $z=0$, $z\sim 1$ and $z\sim
2$ samples. In the bottom row we show S\'ersic index against
rest-frame $u-g$ color for the same galaxies. The S\'ersic indices are
derived from fits to the residual-corrected $H_{160}$-band ($g$-band
for the $z=0$ galaxies) surface brightness profiles. There is a clear
relation between these parameters at all redshifts: star-forming
galaxies have ``diskier'' (lower $n$) profiles and bluer colors than
quiescent galaxies. There is a large variation in the SSFRs, colors
and S\'ersic indices of $z\sim 2$ galaxies, and the spread in these
parameters is of roughly the same order of magnitude in all redshift
bins. We note that the relation shows systematic evolution: the median
SSFR increases with increasing redshift and the median color and
S\'ersic index decrease, from log SSFR = $-11.85^{+1.67}_{-0.22}$
yr$^{-1}$ at $z=0$ to log SSFR = $-8.87^{+0.50}_{-0.58}$ yr$^{-1}$ at
$z\sim 2$, from $u-g_{rest} = 1.62^{+0.38}_{-0.44}$ to $u-g_{rest} =
0.87^{+0.34}_{-0.33}$, and from $n=4.25^{+1.71}_{-2.79}$ to
$n=1.20^{+2.16}_{-0.64}$ (where the errors give the 1-$\sigma$
interval around the median). These results are qualitatively
consistent with the trends derived in e.g., \cite{fra08}.

\begin{figure*}
\epsscale{1.2}
\plotone{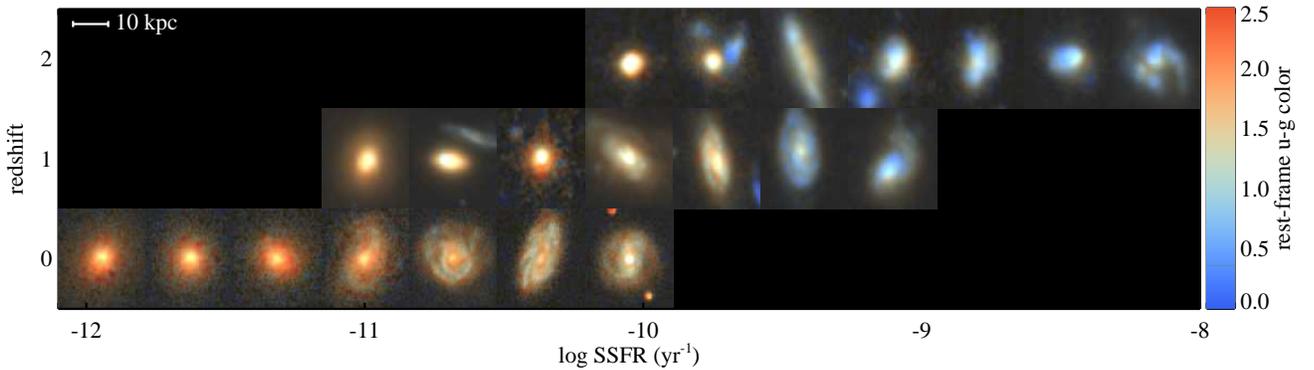}
\caption{Galaxy morphologies from $z\sim 2$ to $z=0$, as a function of
  SSFR. The colors of the galaxies have been obtained by scaling their
  rest-frame $u-g$ colors to the image's RGB space, as indicated by
  the color bar. The range in galaxy structure at $z\sim 2$ is very
  great, and is accompanied by a large range in galaxy colors. There
  is considerable evolution towards redder colors and lower SSFRs from
  $z\sim 2$ to $z=0$. However, signs of a Hubble sequence (i.e., high
  SSFR galaxies are ``diskier'', more extended and bluer than low SSFR
  galaxies) appear to exist at $z\sim 2$.}\label{fig:images}
\end{figure*}

The relations between structure, color and SSFR are further
illustrated in Figure~\ref{fig:images}, which shows rest-frame $u-g$
color images of a selection of galaxies at $z=0$, $z\sim 1$ and $z\sim
2$, as a function of SSFR. From each redshift bin we select seven
galaxies, evenly spaced in log SSFR. This figure very clearly
illustrates the morphological variety present at $z\sim
2$. Additionally, it confirms the relations shown in
Figure~\ref{fig:hubble}: star-forming galaxies are blue and extended,
while quiescent galaxies are red and relatively compact. Thus we find
that the variation in galaxy structure at $z\sim 2$ is as large as at
$z=0$. Furthermore, the systematic relationships between different
structural parameters are very similar between $z=0$ and $z\sim
2$. This, in addition to the lack of evolution in the $u-g_{rest}$
color gradients shown in Figure~\ref{fig:profiles}, suggests that the
underlying mechanisms that give rise to the Hubble sequence at $z=0$
may already be in place at $z\sim 2$.

\section{Discussion}

We have shown that the morphologies of massive $z\sim 2$ galaxies in
the HUDF are complex and varied: from compact, apparently early-type
galaxies to large star forming systems superficially similar to nearby
spirals. Many of these galaxies seem to be composed of multiple
components with large differences in stellar age. We conclude that the
variety in morphologies which is observed at $z = 0$ also exists in
galaxies at $z\sim 2$. This is confirmed by an analysis of the surface
density profiles, which reveals a large range in profile shapes. The
profiles of the large star forming systems are close to exponential,
whereas the profiles of the quiescent systems are more
concentrated. The correlations between morphology and SSFR are similar
at all redshifts between $z=0$ and $z\sim 2$.

This does not mean that morphologies and profiles are static: galaxies
evolve quite strongly between $z =2 $ and $z = 0$. Quiescent galaxies
are much more compact at high redshift, and cannot evolve passively
into low redshift quiescent galaxies (e.g., \citealt{dad05};
\citealt{tru06}; \citealt{tof07}; \citealt{dok08}). A similar size
evolution is required for the star forming galaxies (e.g.,
\citealt{wil10}). We note that due to the evolution in the mass
function of galaxies the $z\sim 2$ galaxies considered here will
almost certainly not evolve into the low-redshift galaxies in this
paper. It would be very informative to examine the morphologies of
number density-limited galaxy samples over this redshift range, in
order to investigate the evolution of the same galaxy population over
time (e.g., \citealt{dok10}).

Additionally, we have derived $u-g_{rest}$ color gradients of our
galaxy sample, and have shown that massive $z\sim 2$ galaxies have
negative color gradients that are comparable to those of low redshift
galaxies in the same mass range. These color gradients may partly be
caused by radial variations in dust extinction, but it is likely that
radial changes in stellar populations play a large role (e.g.,
\citealt{abr99}). This supports a galaxy growth scenario where small
galaxies formed at high redshift grow by accreting material onto their
outer regions. However, more detailed studies are needed in order to
disentangle dust and stellar age effects on the color gradients of
high redshift galaxies.

The results seem to contrast with earlier analyses of galaxies at
$z\sim 2$ in the HDFN (e.g., \citealt{dic00}, \citealt{pap05}). These
authors concluded that all galaxies at this redshift are irregular and
compact, with little difference between the rest-frame UV and
optical. Field-to-field variations may play a role; as shown in
\cite{lab03a} the HDFN contains very few massive high-redshift
galaxies. Indeed, the galaxies studied by these authors have lower
masses ($M_{stellar} \lesssim 3 \times 10^{10} M_\odot$) than the
galaxies we consider, which could result in the difference with the
study presented here. Additionally, We note that recent kinematical
studies of massive $z\sim 2$ galaxies also indicate that many contain
gas with ordered motion (e.g., \citealt{gen08}, \citealt{for09},
\citealt{tac10}). The results presented in this Letter are fully
consistent with those.

Our results raise the question at what redshift the first ``ordered''
galaxies appeared, with structures similar to the Hubble
sequence. There are indications that at redshifts beyond 3 such
galaxies may be much harder to find. Typical high redshift Lyman break
galaxies are very clumpy and irregular (e.g., \citealt{low97}), and
differ significantly in appearance from regular spiral
galaxies. Furthermore, the population of massive galaxies that are
faint in the UV may be very small at redshifts beyond 3 (e.g.,
\citealt{bra07}), at least in the mass regime that is considered in
this paper (e.g., \citealt{mar10}). With current observational
capabilities we are severely limited in studying the rest-frame
optical properties of galaxies at redshifts beyond $z\sim 3.5$, due to
rest-frame optical emission moving redward of the observers' $K$
band. With improved capabilities it may become possible to study the
red massive galaxy population at these redshifts. Several candidates
have been found at $z > 5$ with significant Balmer discontinuities
(\citealt{eyl05}; \citealt{mob05}; \citealt{wik08}; \citealt{ric11}),
and many more are speculated to exist. These galaxies could very well
be the centers of multi-component galaxies at redshifts between $z =
3$ and $z = 5$. 

\acknowledgments

We thank the Leids Kerkhoven Bosscha foundation for providing travel
support. We thank the Lorentz Center for hosting workshops during
which this paper was written.  We thank Joop Schaye, Phil Hopkins,
Lars Hernquist and Rachel Somerville for discussions.  Support from
NASA grant HST-GO-10808.01-A is gratefully acknowledged.

{\it Facilities:} \facility{HST (ACS, WFC3)}, \facility{VLT (ISAAC)}.

\end{document}